\documentclass[prl,aps,twocolumn,preprintnumbers,10pt]{revtex4}

\usepackage{amssymb,amsmath,amsfonts,amsbsy,epsfig}

\def\mathbi#1{\textbf{\em #1}}

\newcommand{\mpl}{m_{\rm Pl}}
\newcommand{\fnl}{f_{\rm NL}}
\newcommand{\calO}{{\cal O}}
\newcommand{\calP}{{\cal P}}
\newcommand{\calR}{{\cal R}}

\begin{document}

\preprint{CERN-PH-TH/2012-315}
\preprint{APCTP-Pre2012-016}

\title{Correlating features in the primordial spectra}
\date{\today}
\author{
Ana Ach\'ucarro$^{a,b}$, Jinn-Ouk Gong$^{c,d,e}$, Gonzalo A. Palma$^{f}$ and Subodh P. Patil$^{c}$
}
\affiliation{
\vspace{5pt}
$^{a}$Instituut-Lorentz for Theoretical Physics, Universiteit Leiden, 2333 CA Leiden, The Netherlands
\\
$^{b}$Department of Theoretical Physics, University of the Basque Country, 48080 Bilbao, Spain
\\
$^{c}$Theory Division, CERN, CH-1211 Gen\`{e}ve 23, Switzerland
\\
$^{d}$Asia Pacific Center for Theoretical Physics, Pohang 790-784, Korea 
\\
$^{e}$Department of Physics, Postech, Pohang 790-784, Korea
\\
$^{f}$Physics Department, FCFM, Universidad de Chile, Blanco Encalada 2008, Santiago, Chile
}

\begin{abstract} 

Heavy fields coupled to the inflaton reduce the speed of sound in the effective theory of the adiabatic mode each time the background inflationary trajectory deviates from a geodesic. This can result in features in the primordial spectra. We compute the corresponding bispectrum and show that if a varying speed of sound induces features in the power spectrum, the change in the bispectrum is given by a simple formula involving the change in the power spectrum and its derivatives. In this manner, we provide a uniquely discriminable signature of a varying sound speed for the adiabatic mode during inflation that indicates the influence of heavy fields. We find that features in the bispectrum peak in the equilateral limit and, in particular, in the squeezed limit we find considerable enhancement entirely consistent with the single field consistency relation. From the perspective of the underlying effective theory, our results generalize to a wide variety of inflationary models where features are sourced by the time variation of background quantities. A positive detection of such correlated features would be unambiguous proof of the inflaton's nature as a single light scalar degree of freedom embedded in a theory that is UV completable.
\end{abstract}

\maketitle 


Although no significant evidence for features in the primordial power spectrum has been observed to date~\cite{features}, a 10\% modulation of power remains entirely consistent with direct reconstruction~\cite{reconstruction} of that region of the power spectrum accessible to present cosmic microwave background (CMB) observations~\cite{wmap}. Further improvement in our knowledge of the matter spectrum beyond the CMB scales from future large scale structure surveys~\cite{future_lss} ought to furnish far superior statistics that may yet allow us to determine if there are any other scales present in the primordial power spectrum other than that which sets its amplitude~\cite{lss-probes}.

Inflationary models that generate features have been considered in the literature for various theoretical motivations. Examples include explaining outliers to the best fit $\Lambda$CDM models in the CMB~\cite{outliers} (see however~\cite{ACT}), probing and constraining couplings of the inflaton to other fields through particle production~\cite{pp}, probing modified vacuum structure \cite{vac}, interrupted slow-roll \cite{interrupted} and perhaps even signatures of inflation's embedding in a UV complete theory \cite{resonance}.

The perspective of this article is informed by recent findings that features induced by reduced and varying speeds of sound  $c_s$ of the adiabatic mode might be an inevitable consequence of field excursions during inflation~\cite{Achucarro:2010jv,Achucarro:2010da,Cespedes:2012hu,Achucarro:2012sm,Achucarro:2012yr, Gao:2012uq}, and the surprisingly large role played by heavy fields in the dynamics of light fields (first noticed in~\cite{Tolley:2009fg}, additionally elaborated upon in~\cite{Burgess:2012dz}). Specifically, heavy fields with masses much larger than the Hubble scale can influence the dynamics of the adiabatic mode in a manner that is entirely consistent with the persistence of slow-roll~\cite{Achucarro:2010da,Cespedes:2012hu}, the validity of an effectively single field description and the decoupling of the true high and low energy modes of the system~\cite{Achucarro:2012yr}. This effective theory remains weakly coupled up to the cutoff that defines it~\cite{weaklycoupled}. At low energies, the effective theory of the adiabatic mode is an operator expansion parametrized entirely by $c_s$~\cite{Achucarro:2012sm}, which can in fact be a relatively rapidly varying function, subject to the adiabaticity condition discussed below.

Here we show that the generic consequence of a varying $c_s$ would be to imprint features in the power spectrum and enhance the bispectrum for all shapes, peaking in the equilateral but also enhancing other configurations (see~\cite{var-cs-features} for similar studies). Our main result is summarized by (\ref{fnlgen}), where $f^\triangle_{\rm NL}$ parametrizes the bispectrum as in (\ref{fNL-B}), in a particular configuration $\triangle$ of the three wavevectors. There, the non-trivial scale dependence of $f^\triangle_{\rm NL}$ is directly correlated to the features in the power spectrum parametrized by $\Delta\calP_\calR/\calP_\calR$, while the $c^{\triangle}$-coefficients depend only on the shape of the specific configuration $\triangle$. This enhancement of the primordial bispectrum in combination with correlated features in the power spectrum could be readily discernible through future observations. 


Our starting point is the effective action for the adiabatic mode. The quadratic action $S_2$ obtained by integrating out a heavy field in a two-field model is given by 
\begin{equation}\label{S2eff}
S_2 = \mpl^2 \! \int \!  \frac{a^3\epsilon}{c_s^2} \left[ \dot\calR^2 - c_s^2\frac{(\nabla\calR)^2}{a^2}\right] \, ,
\end{equation}
where $a$ is the scale factor, $H = \dot a/a$ and $\epsilon = - \dot H / H^2$. Here, $c_s$ is the speed of sound of the adiabatic fluctuations given by~\cite{Achucarro:2012sm}
\begin{equation}
\frac{1}{c_s^2}  \equiv  1 + \frac{4\dot\theta^2}{M_\text{eff}^2} \, ,
\end{equation}
where $M^2_{\rm eff} = M^2 - \dot\theta^2$ is the effective mass of the normal (isocurvature)  fluctuations off the background trajectory, $\dot\theta$ is the angular velocity of the background in field space~\cite{Achucarro:2010jv,Achucarro:2012sm}, and $M^2$ is its mass$^2$ in the absence of any turns. A particular situation of interest is when the background trajectory departs off the adiabatic minimum of the potential as inflation progresses due to `bends' in field space, resulting in transient reductions in the speed of sound~\footnote{Subject to the adiabatic condition that $|\dot c_s| \ll M|1-c_s^{2}|$~\cite{Cespedes:2012hu,Achucarro:2012yr}, which ensures that no heavy quanta are produced even as the background trajectory deviates off the adiabatic minimum.}~\cite{Achucarro:2010jv,Achucarro:2010da}. Presuming that this reduction in the speed of sound is sufficiently small ($|1-c_s^{-2}| \ll 1$) we conveniently rewrite the quadratic part of the action (\ref{S2eff}) as
\begin{equation}\label{actiondeclassification}
S_2 = S_{2,\text{free}} + \underbrace{ \int d^4x a^3\epsilon\mpl^2 \left( \frac{1}{c_s^2} - 1 \right)\dot\calR^2 }_{\equiv S_{2,\text{int}}} \, ,
\end{equation}
where $S_{2,\text{free}}$ is nothing but $S_2$ with $c_s = 1$. With this splitting of the quadratic action, $S_{2,\text{int}}$ may be considered as a perturbation to $S_{2,\text{free}}$. Thus one can readily evaluate the corrections to the power spectrum and the bispectrum induced by a varying $c_s$. Denoting~\cite{Seery:2005wm}
\begin{equation}
u \equiv 1 - \frac{1}{c_s^2} \, ,
\end{equation}
the change in the power spectrum generated by changes in the speed of sound, to first order in $u$, is given by~\cite{bispectrum}
\begin{equation}
\label{ps}
\frac{\Delta\calP_\calR}{\calP_\calR}(k) =  k  \int_{- \infty }^0 d  \tau  \, u(\tau) \sin\left( 2 k \tau \right) \, ,
\end{equation}
where $\calP_\calR = H^2/(8\pi^2\epsilon\mpl^2)$ is the featureless flat power spectrum and $\tau$ is the conformal time. One immediately sees how features in the power spectrum are generated by a varying $c_s$. We also note that constant, reduced speeds of sound renormalize the power spectrum without imparting any new scale dependence. 

One can also compute the leading contributions to the bispectrum $\Delta B_\calR$ due to a non-vanishing $u$, as~\cite{bispectrum}
\begin{widetext}
\vspace{-0.5cm}
\begin{align}\label{bispectrum}
\Delta B_\calR(\mathbi{k}_1,\mathbi{k}_2,\mathbi{k}_3) & = 2\Re \left\{ 2i \widehat{\calR}_{k_1}(0)\widehat{\calR}_{k_2}(0)\widehat{\calR}_{k_3}(0)  \left[3 \epsilon \frac{\mpl^2}{H^2} \int_{-\infty}^0 \frac{d\tau}{\tau^2} u (\tau) \frac{d\widehat{\calR}_{k_1}^*(\tau)}{d\tau} \frac{d\widehat{\calR}_{k_2}^*(\tau)}{d\tau} \widehat{\calR}_{k_3}^*(\tau) + \text{2 perm} \right.\right. \\
& \left.\left. \hspace{3.5cm} + \epsilon\frac{\mpl^2}{H^2} \left( \mathbi{k}_1\cdot\mathbi{k}_2 + \text{2 perm} \right) \int_{-\infty}^0 \frac{d\tau}{\tau^2} \left( u - \tau \frac{du}{d\tau} \right) \widehat{\calR}_{k_1}^*(\tau)\widehat{\calR}_{k_2}^*(\tau)\widehat{\calR}_{k_3}^*(\tau) \right] \right\}\nonumber \, ,
\end{align}
\end{widetext}
where the mode function solution $\widehat{\calR}_k(\tau)$ is given by
\begin{equation}\label{modesolution}
\widehat{\calR}_k(\tau) = \frac{iH}{\sqrt{4\epsilon k^3}\mpl} \left( 1 + ik\tau \right) e^{-ik\tau} \, .
\end{equation}
In the above, we work to leading order in slow-roll parameters and to linear order in $u$. The complete expression for the bispectrum is of the form $B = B_0 + \Delta B$, where $B_0$ represents the leading non-zero contributions when $c_s=1$ and is of $\calO(\epsilon^2)$. $\Delta B$ dominates whenever $u$'s maximum value $|u|_{\rm max}$ is larger than $\mathcal O (\epsilon)$. From (\ref{ps}) we see that $|u|_{\rm max}$ of $\mathcal O(10^{-1})$ translates into features in the power spectrum of $\calO(10) \%$, which is reasonable for the scales accessible to the CMB to this level. Thus, with slow-roll parameters no bigger than $\mathcal O(10^{-2})$, we obtain that $\Delta B$ would become the dominant contribution to the bispectrum.

\begin{figure*}[t]
 \begin{center}
  \includegraphics[width=15.7cm]{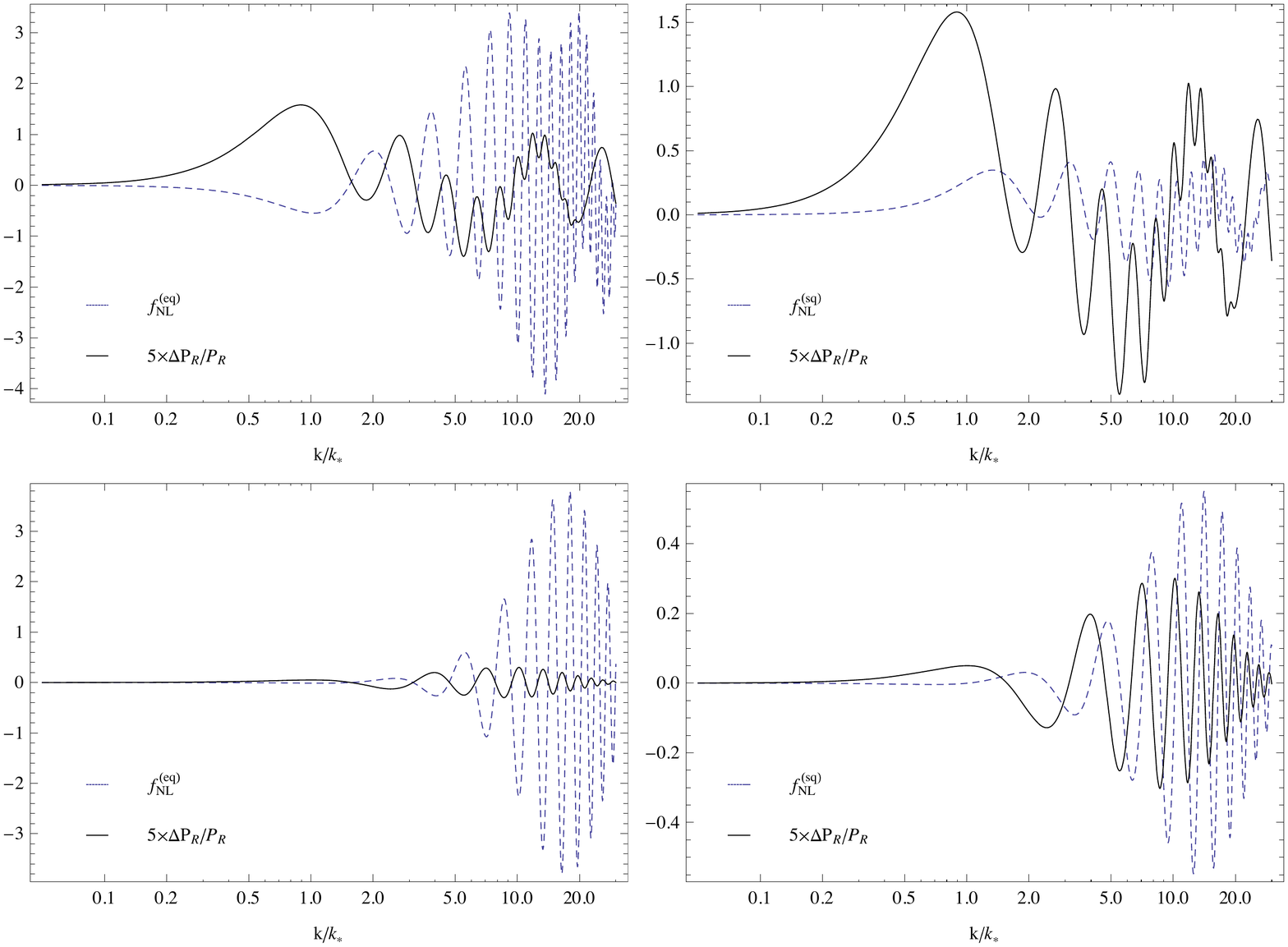}
 \end{center}
\vspace{-2.5em}
\caption{$\Delta\calP_\calR/\calP_\calR$ (solid line) versus $f_\mathrm{NL}^\mathrm{(eq)}$ (equilateral, dotted line, left panel) and $f_\mathrm{NL}^\mathrm{(sq)}$ (squeezed, dotted line, right panel) for $u = -u_\mathrm{max} \left\{ \tanh\left[(N-N_i)/\delta{N}\right] - \tanh\left[(N-N_f)/\delta{N}\right] \right\}$ with $u_\mathrm{max}=1/12$, $N_f-N_i=2$ and $\delta{N}=0.05$ (top panel), and $u = -u_\mathrm{max}/\cosh^4\left[(N-N_\star)/\Delta{N}\right]$ with $u_\mathrm{max}=1/12$ and $\Delta{N}=0.2$. $N=-\log|\tau|$ is the number of $e$-folds, $\star$ denotes a convenient reference, and here the central moment of the turn, and $N_\star$ and $k_\star$ are the corresponding values.}
 \label{fig:BPcorr}
\end{figure*}

Taken separately, one immediately sees from (\ref{ps}) and (\ref{bispectrum}) how the changes in $c_s$ parametrized by $u$ source scale dependence in both the power spectrum and the bispectrum. However to highlight their correlation in a manner that should serve as a useful discriminant in probing data, it is useful to invert $u$ to linear order in terms of the change in the power spectrum as
\begin{equation}
\widetilde{u}(\tau) = \frac{2i}{\pi} \int^\infty_{-\infty} \frac{d k}{k} \frac{\Delta\calP_\calR}{\calP_\calR}(k) e^{-2i k \tau} \, ,
\end{equation}
where $\widetilde{u}$ is defined as the odd extension of $u$ over the real line. Substituting this into (\ref{bispectrum}) allows us to infer the leading contribution to the bispectrum as
\begin{widetext}
\begin{align}\label{bispectrumf}
\Delta B_\calR(\mathbi{k}_1,\mathbi{k}_2,\mathbi{k}_3) = & \frac{(2\pi)^4\calP_\calR^2}{(k_1k_2k_3)^2} \left\{ -\frac{3}{2} \frac{k_1k_2}{k_3} \left[ \frac{1}{2k} \left( 1 + \frac{k_3}{2k} \right) \frac{\Delta\calP_\calR}{\calP_\calR}( k) - \frac{k_3}{4k^2} \frac{d}{d\log k} \left( \frac{\Delta\calP_\calR}{\calP_\calR} \right) \right] + \text{2 perm} \right.\\
& \hspace{1.7cm} + \frac{1}{4} \frac{k_1^2+k_2^2+k_3^2}{k_1k_2k_3} \left[ \frac{1}{2k} \left( 4k^2 - k_1k_2 - k_2k_3 - k_3k_2 - \frac{k_1k_2k_3}{2k} \right) \frac{\Delta\calP_\calR}{\calP_\calR}( k) \right.
\nonumber\\
& \left.\left. \hspace{3.3cm} - \frac{k_1k_2+k_2k_3+k_3k_1}{2k} \frac{d}{d\log k} \left( \frac{\Delta\calP_\calR}{\calP_\calR} \right) + \frac{k_1k_2k_3}{4k^2} \frac{d^2}{d\log k^2} \left( \frac{\Delta\calP_\calR}{\calP_\calR} \right) \right] \right\}\biggl|_{k = (k_1 + k_2 + k_3)/2}\nonumber \, ,
\end{align}
where $k_i \equiv |\mathbi{k}_i|$, and $\calP_\calR, \ \Delta \calP_\calR$ are always evaluated at $k \equiv (k_1 + k_2 +
k_3)/2$~\cite{bispectrum}. This is the result we wish to highlight:
features in the power spectrum translate directly into correlated
features in the bispectrum, with the precise nature of the k
correlation depending on the configuration we look at. We may define a
dimensionless shape function, with the $\fnl$ ansatz in mind, as
\begin{equation}\label{fNL-B}
f_\mathrm{NL}^\triangle(k_1,k_2,k_3) \equiv \frac{10}{3} \frac{k_1k_2k_3}{k_1^3+k_2^3+k_3^3} \frac{(k_1k_2k_3)^2 \Delta B_\calR}{(2\pi)^4\calP_\calR^2} \, ,
\end{equation}
where the shape superscript on the left hand side indicates that it is a function of particular configurations of the three wavevectors. Evaluating the above in certain interesting limits, we find
\begin{equation}\label{fNLs}
f_\mathrm{NL}^\triangle = \left\{
\begin{array}{ll}
\dfrac{5}{54} \left[ -7\dfrac{\Delta\calP_\calR}{\calP_\calR} - 3 \dfrac{d}{d\log k} \left( \dfrac{\Delta\calP_\calR}{\calP_\calR} \right) + \dfrac{d^2}{d\log k^2} \left( \dfrac{\Delta\calP_\calR}{\calP_\calR} \right) \right]\biggl|_{k = (k_1 + k_2 + k_3)/2} & \left( \dfrac{k_2}{k_1}=\dfrac{k_3}{k_1}=1 \text{: equilateral} \right)
\\
-\dfrac{5}{12} \dfrac{d}{d\log k} \left( \dfrac{\Delta\calP_\calR}{\calP_\calR} \right)\biggl|_{k = (k_1 + k_2 + k_3)/2} & \left( \dfrac{k_2}{k_1}=1, \dfrac{k_3}{k_1}\to0 \text{: squeezed} \right)
\\
\dfrac{1}{8} \left[ -\dfrac{\Delta\calP_\calR}{\calP_\calR} - \dfrac{5}{2} \dfrac{d}{d\log k} \left( \dfrac{\Delta\calP_\calR}{\calP_\calR} \right) + \dfrac{1}{2} \dfrac{d^2}{d\log k^2} \left( \dfrac{\Delta\calP_\calR}{\calP_\calR} \right) \right]\biggl|_{k = (k_1 + k_2 + k_3)/2} & \left( \dfrac{k_2}{k_1}=1, \dfrac{k_3}{k_1}=2 \text{: folded} \right)
\end{array}
\right. .
\end{equation}
\end{widetext}
We notice that the squeezed limit is nothing other than a re-expression of the single field consistency relation~\cite{consistency} 
\begin{equation}\label{consistency}
B_\calR(\mathbi{k}_1,\mathbi{k}_2,\mathbi{k}_3) \underset{k_1\to0}{\longrightarrow} - P_\calR(k_1)P_\calR(k_3)\frac{d\log P_\calR(k_3)}{d\log k_3} \, ,
\end{equation}
with $P_\calR(k) = 2\pi^2\calP_\calR/k^3$. Thus $f^{\rm (sq)}_{\rm NL}$ may approach values of order unity, even though slow-roll is operative throughout. This is because the spectral index receives contributions that go like $s \equiv \dot c_s/(Hc_s)$~\cite{generalsinglefield-running} which can approach order unity consistent with slow-roll and the validity of the single field approximation~\cite{Achucarro:2010da,Cespedes:2012hu,Achucarro:2012sm,Achucarro:2012yr}. From (\ref{fNLs}) we infer that in general, for any configuration one might look at, the bispectrum is sourced by features in the power spectrum as
\begin{equation}\label{fnlgen}
f^\triangle_{\rm NL} \sim c^\triangle_0(\mathbi{k})\frac{\Delta\calP_\calR}{\calP_\calR} + c^\triangle_1(\mathbi{k})\left(\frac{\Delta\calP_\calR}{\calP_\calR}\right)' + c^\triangle_2(\mathbi{k})\left(\frac{\Delta\calP_\calR}{\calP_\calR}\right)'' \, ,
\end{equation}
with a prime denoting a logarithmic derivative and the coefficients $c^\triangle_i$ depending only on the shape of the configuration we look at, and with all information about the variation of $c_s$ encoded in $\Delta\calP_\calR/\calP_\calR$. In Figure~\ref{fig:BPcorr} we plot the changes in the power spectrum against the bispectrum for a few prototypical changes in the speed of sound that model different varieties of turns in the background trajectory. As evident from the plots, the shape function in each limit (\ref{fNLs}) exhibits features in accordance with modulations in the power spectrum and its derivatives with the amplitude of the equilateral $f_\mathrm{NL}^\mathrm{(eq)}$ being 
as high as $\sim 3$ for the parameters we consider if we take $|u|_{\rm max} \sim 10^{-1}$. More interestingly, the squeezed configuration, in addition to being $\pi/2$ out of phase with the modulations of the power spectrum as expected from the consistency relation (\ref{consistency}), attains a peak value $f_\mathrm{NL}^\mathrm{(sq)} \sim 0.5$. Although this is very unlikely to be detectable in the CMB, the prospects for measuring $\fnl$ to a precision of $\calO(1)$ at smaller scales stands to improve with future observations~\cite{future_lss}.

From the perspective of the effective field theory of inflation of~\cite{Cheung:2007st}, the cubic order action is parametrized by coefficients $M^4_2$ and $M^4_3$ in such a way that $c_s$ is determined by $M^4_2$. On general grounds~\cite{Senatore:2009gt} one expects $M_3$ to be of the form $M_3^4 \sim \left(1 - c_s^{-2} \right)M_2^4$, with the precise relation encoding properties of the parent theory from which the effective theory descends. For example, $M^4_3/M^4_2 = 3(1- c_s^{-2})/2$ for DBI inflation, whereas $M^4_3/M^4_2 = 3(1- c_s^{-2})/4$ when one heavy field has been integrated out in a two-field theory~\cite{Achucarro:2012sm}. Because $M_3^4 \sim \left(1 - c_s^{-2} \right)M_2^4$ implies terms of $\calO(u^2)$ in (\ref{bispectrum}), we see that (\ref{bispectrumf}) and (\ref{fNLs}) are representative of a wide class of single field models of inflation, beyond the specific case in which adiabatic modes interact with heavy fields. In spite of these details, our results show that one will always arrive at the general expression (\ref{fnlgen}) where the coefficients $c^\triangle_i$ depend on the details of the parent theory as well as the shape of the configuration we are interested in, incorporating for example the setups studied in \cite{feat-pot}.


The prospects for observing a non-trivial scale dependence in the primordial bispectrum are enhanced if they are correlated at commensurate comoving scales with features in the power spectrum. In this article, we illustrate a context in which this occurs naturally-- when the dynamics of the adiabatic mode are influenced by heavy fields in such a way as to transiently reduce $c_s$ at various points along the inflationary trajectory, consistent with the persistence of slow-roll and the validity of the effective single field description \cite{Achucarro:2010da,Achucarro:2010jv,Achucarro:2012sm,Achucarro:2012yr}. In this way, we offer a uniquely discriminable signature of the effect of higher dimensional operators that become relevant during inflation, the positive observation of which would allow us to infer properties of the parent theory in which inflation is embedded.

\begin{acknowledgements}

We wish to thank Thorsten Battefeld, Vincent Desjacques and Jan Hamann for useful discussions. 
AA is supported by funds from the Netherlands Foundation for Fundamental Research on Matter (F.O.M), Basque Government grant IT559-10, the Spanish Ministry of Science and Technology grant FPA2009-10612 and Consolider-ingenio programme CDS2007-00042.
JG is supported by a Korean-CERN fellowship, and acknowledges the Max-Planck-Gesellschaft (MPG), the Korea Ministry of Education, Science and Technology, Gyeongsangbuk-Do and Pohang City for the support of the Independent Junior Research Group at the Asia Pacific Center for Theoretical Physics.
GAP is supported by Conicyt under the Anillo project ACT1122 and the Fondecyt Regular Project No. 1130777.
SPP is supported by a Marie Curie Intra-European Fellowship of the European Community's 7th Framework Programme under contract number PIEF-GA-2011-302817.
\end{acknowledgements}

\end{document}